\documentclass[manuscript,screen,nonacm]{acmart}

\AtBeginDocument{%
  }

\setcopyright{none}
\acmYear{2026}
\copyrightyear{2026}

\usepackage{booktabs}
\usepackage{tabularx}
\usepackage{listings}
\usepackage{xcolor}

\begin{document}

\title{Orchestrated Reality: From Role-Play to Living, Playable Game
Worlds}
\subtitle{LLM-Driven World Simulation as a Parameterized-Action POMDP}

\author{Yuhang Huang}
\authornote{Yuhang Huang and Chenmiao Li contributed equally to this work.}
\affiliation{%
  \institution{The University of Tokyo}
  \country{Japan}
}

\author{Chenmiao Li}
\authornotemark[1]
\affiliation{%
  \institution{The University of Tokyo}
  \country{Japan}
}

\author{Chaowei Fang}
\affiliation{%
  \institution{Individual Researcher}
  \country{Japan}
}

\renewcommand{\shortauthors}{Huang, Li, and Fang}

\begin{abstract}
Many games rely on storytelling combined with systems that track
levelling, NPC behaviour, and consequence simulation; bridging
tightly-authored narrative with deeply-simulated worlds---most acute in
sandbox and open-world settings---has been prohibitively expensive.
LLM-driven worlds open a new path: a single harness can coordinate
numerical state, narrative voice, storytelling pacing, and rule logic
together. Realising this requires the LLM system to sustain a
persistent world (who is where, what has just happened, what is
currently true), which today's deployed systems do not: the narrative
voice \emph{asserts} state in free prose without any validated
representation, so a fully autonomous game engine remains infeasible. We treat this as an
architectural choice, not a limitation of language models, and report
work in progress on a framework---\emph{orchestrated reality}---that
makes the world a canonical object owned by a singleton orchestration
agent analogous to the tabletop-RPG \textbf{Game Master (GM)}. We
formalise an LLM-driven game world for a human player as a
\textbf{Parameterized-Action POMDP}~\cite{kaelbling1998planning,
masson2016parameterized}: state is a tree of canonical JSON entities,
actions decompose as $a=(k, x_k)$ (a discrete intent kind plus
structured JSON parameters), the agent observes only a narrative
projection $o=O(s)$ of state, and the transition kernel $F$ is an
LLM-driven \textbf{Plan--Diff--Validate--Apply (PDVA)} pipeline that
commits schema-validated, content-hashed JSON deltas. We give the
formal model, a JSON-state example, a worked single-turn example, and
a catalogue of 15 illustrative incidents drawn from a real deployment
showing the framework in action. Empirical validation through a planned human
player study---together with multi-NPC concurrent agency and
deployment as an RL environment---is situated as future work.
\end{abstract}



\maketitle

\section{Introduction}

Many games rely on storytelling, or at minimum on systems that track
levelling and status, narrate non-player characters, and resolve
player actions into simulated outcomes. In more open-ended settings
such as sandbox and open-world games, the storytelling and memory
burden needed to make the world feel real becomes considerably more
complex. Today's LLM-based systems frequently fail to maintain a
coherent world state and a reasonable narrative system, which is why
LLM-based narrative works only for character role-play scenarios and
has not yet emerged into real computer games. They lose track of who
is where, what has just happened, and what is currently true, which
makes it infeasible to build a fully autonomous game engine without a
human-in-the-loop update mechanism. The market for LLM-driven games
is real; however, most current exploration focuses on the game-coding
environment and on generating 3D art and asset environments, rather
than on world state and narrative orchestration. If
we can move from character-based narrative generation to
game-story-orchestration agents---analogous to \textbf{Game Masters
(GMs)} in tabletop role-playing games (TRPGs)---such agentic systems
could become the foundation of next-generation games. This is the
move from a character that speaks to \emph{orchestrated reality}.

We treat this as an architectural problem, not a language-model
limitation. In essentially every deployed or studied system, the
narrative voice \textbf{asserts world state in free prose} rather than
reading and writing a persistent, validated representation. This
produces three recurring failures. (1)~\emph{Statelessness}: when a
session ends the world ceases to exist; ``memory'' is only
context-window proximity. (2)~\emph{Unvalidated writes}: the model
can assert any world change, so state silently drifts and accumulates
contradictions. (3)~\emph{Monolithic agency}: one prompt simultaneously
narrates the environment, voices every character, and adjudicates
rules, conflating roles that should be separated.

We report work in progress on a framework that makes the world the
canonical object. The world is a tree of \textbf{JSON entities} on
disk---this is the state. Each turn the system advances state through
a \textbf{Plan--Diff--Validate--Apply (PDVA)} pipeline that proposes,
schema- and permission-validates, and atomically commits JSON deltas;
this is the transition kernel. The agent (the player today; an RL
policy or sub-agent in future work) emits \textbf{parameterized
actions} $a=(k, x_k)$---a discrete intent kind $k$ plus structured JSON
parameters $x_k$, and observes only a narrative projection $o=O(s)$
of canonical state. We formalise this loop as a
\textbf{Parameterized-Action POMDP}, combining the partially
observable setting of Kaelbling, Littman, and
Cassandra~\cite{kaelbling1998planning} with the parameterized-action
setting of Masson, Ranchod, and Konidaris~\cite{masson2016parameterized}.

\paragraph{Contributions (work in progress).}
\begin{enumerate}
  \item A formalisation of LLM-driven game worlds as a
  \emph{Parameterized-Action POMDP}, with a single world-agent serving
  as both the transition kernel ($F$, via PDVA) and the observation
  kernel ($O$, via narration).
  \item A \emph{JSON world-state} model that makes the world
  inspectable, branchable, and replayable by content hash---a concrete
  answer to ``what would it mean to actually model the world.''
  \item A catalogue of \emph{15 auditable incident classes} drawn from a real
  deployment that illustrate the framework's three mechanisms---multi-agent
  orchestration, JSON-forced memory injection, and JSON-flow state
  commits---in action, demonstrating that the framework runs in practice
  on real play history.
\end{enumerate}

\section{Related Work}

\textbf{Markov decision processes for game environments.} Our
formalism joins two MDP threads: partial observability
(POMDPs~\cite{kaelbling1998planning}) and parameterized actions
(PAMDPs~\cite{masson2016parameterized}). Multi-agent extensions go
through Markov games~\cite{littman1994markov}, which we reserve for
future work in which NPCs become concurrent actors. In ML, learned
\emph{world models}~\cite{ha2018world} have been used as RL
environments; we propose a complementary line where the world model is
made of structured JSON and the transition kernel is an LLM-driven
validation pipeline, not a learned neural network.

\textbf{AI game masters and agentic orchestration.} Comparisons of
static versus agentic AI game masters~\cite{agenticgm2025} report
immersion and curiosity gains for agentic orchestration but treat the
world's underlying state as implicit. Studies of player trust under
LLM-driven NPCs~\cite{yin2024trust} measure experiential effects of
inconsistency, not the underlying architecture of memory and state. We
address that layer: not the GM's prose choices, but the validated
world model under them.

\textbf{LLM-driven worlds and LLM agents.} Generative
Agents~\cite{park2023generative} build an \emph{LLM-driven world}
without a human player; Voyager~\cite{wang2023voyager} and
MineDojo~\cite{fan2022minedojo} place LLMs as policies in a fixed
Minecraft world; RoleLLM~\cite{wang2024rolellm} and
Character-LLM~\cite{shao2023characterllm} target persona consistency
inside an agent. We target the missing intersection: an
\emph{LLM-driven world for a human player}, formalised so the world
responds turn-by-turn to parameterized actions.

\textbf{Story generation with agents.}
Dramatron~\cite{mirowski2023dramatron} co-writes scripts with humans
through an agentic prompt structure; Agents'
Room~\cite{huot2025agentsroom} generates narratives through multi-step
collaboration. These pipelines produce text; we provide the runtime
world model that any such pipeline can read from and write into
through a validated transition kernel.

\textbf{Multi-agent LLM frameworks.}
AutoGen~\cite{wu2023autogen}, MetaGPT~\cite{hong2023metagpt}, and
CAMEL~\cite{li2023camel} provide general multi-agent orchestration but
no game-specific agent contracts, no schema-validated mutation
pipeline, and no event-sourced state with hash-identified replay. We
bring those systems disciplines---deny-first permissions, atomic
transactions, content-addressed audit logs---to a game-runtime
setting.

\section{Formalism: World as a Parameterized-Action POMDP}

\subsection{The world is a tree of JSON entities}

Many agent systems use the filesystem incidentally---scratchpads, RAG
corpora, run logs. Our position is stronger: we use JSON files as the
\emph{canonical state of the world}, where ``world'' foregrounds
\textbf{narrative, characters, and cultural fact} as first-class
state rather than only spatial or asset layout. Every entity that
matters---a town, an NPC with persona and relationships, a quest's
history, a faction's standing, the active narrative thread, the
player profile, the run state itself---is a JSON document under a
typed schema; the world $s\in \mathcal{S}$ is the tree of these
documents on disk. State is
therefore inspectable, diffable, and addressable by content hash
\emph{before any model is invoked}. As a concrete sketch:

\begin{lstlisting}[language=]
# game/meta/run_state.json
{ "location": {"scope": "town", "node_id": "T001", "subnode_id": "gate_north"},
  "time": {"day": 1, "clock": "08:30"}, "turn_count": 0 }

# game/towns/T001/town.json
{ "id": "T001", "name": "Stone Ford", "region": "Northvale",
  "entry_points": ["gate_north", "dock"],
  "laws":  {"weapon_policy": "peace_bonded_in_townhall"},
  "state": {"alert": 0, "recent_events": []} }

# game/player/profile.json
{ "name": "Player", "background": "wanderer", "class": "fighter" }
\end{lstlisting}

Each turn reads a small slice of this tree, proposes a delta, and
commits a new tree---never replaces ``the world'' with a text blob.

\subsection{The Parameterized-Action POMDP}

We model an LLM-driven game world as the tuple
\begin{equation}
  \mathcal{M} = (\mathcal{S},\ \mathcal{A},\ \Omega,\ O,\ F,\ \rho_0).
\end{equation}
$\mathcal{S}$ is the space of JSON state trees introduced in \S3.1:
each $s\in\mathcal{S}$ is a tree of typed JSON documents on disk under
a fixed schema, fully serialisable and addressable by content hash
$h(s)$. Actions are \emph{parameterized} in the sense of Masson et
al.~\cite{masson2016parameterized}:
\begin{equation}
  \mathcal{A} = \{(k,\ x_k)\mid k\in\mathcal{A}_d,\ x_k\in\mathcal{X}_k\},
\end{equation}
where $\mathcal{A}_d$ is the finite set of intent kinds declared by the
world schema (e.g.\ \texttt{move}, \texttt{speak}, \texttt{give},
\texttt{inspect}) and, for each $k$, $\mathcal{X}_k$ is the space of
$k$-specific structured JSON parameters. $\Omega$ is the space of player-facing
observations---natural-language narration plus surfaced status
(location, time, inventory). The agent does not observe $s_t$
directly; it receives $o_t \sim O(s_t)$, where
$O:\mathcal{S}\to\Delta(\Omega)$ is the narration kernel emitted by
the world-agent---this is the POMDP~\cite{kaelbling1998planning} side
of the formalism. The transition kernel
\begin{equation}
  F: \mathcal{S}\times\mathcal{A}\to\Delta(\mathcal{S})
\end{equation}
is the \textbf{Plan--Diff--Validate--Apply (PDVA)} pipeline. Finally,
$\rho_0\in\Delta(\mathcal{S})$ is the initial-state distribution: a
scenario template (e.g.\ ``the player arrives at the north gate, day~1,
08:30'') instantiated into the JSON tree of \S3.1 as the starting
$s_0\sim\rho_0$. Given
$s_t$ and $a_t=(k_t, x_{k_t})$:

\begin{itemize}
  \item \textbf{Plan.} Build a bounded context
  $p=\mathrm{Context}(s_t, a_t)$ from the canonical state slice
  relevant to $a_t$.
  \item \textbf{Diff.} A model returns
  $r=(\mathrm{narrative},\,\Delta_t)$: the prose for the next
  observation $o_{t+1}$ plus a structured proposed mutation
  $\Delta_t=\{(\mathrm{path}_j,\mathrm{op}_j,\mathrm{value}_j)\}$.
  \item \textbf{Validate.} Each $\delta_j\in\Delta_t$ must pass
  \begin{equation}
    \mathrm{Valid}(\delta_j)=
    \mathrm{SchemaOK}(\delta_j)\,\wedge\,
    \mathrm{PermOK}(\delta_j,c)\,\wedge\,
    \mathrm{RuleOK}(\delta_j,\mathcal{R}),
  \end{equation}
  i.e.\ matches the JSON schema, lies inside the proposing actor's
  permission scope $c$, and satisfies game rules $\mathcal{R}$.
  \item \textbf{Apply.} Valid changes commit atomically and the
  resulting state is content-hashed:
  \begin{equation}
    s_{t+1}=\mathrm{Apply}\!\big(s_t,\,\{\delta_j:\mathrm{Valid}(\delta_j)\}\big),
    \quad h(s_{t+1})=\mathrm{SHA256}(\mathrm{Serialize}(s_{t+1})).
  \end{equation}
\end{itemize}

\paragraph{Markov property by construction.} A raw LLM chat has no
Markov structure: the ``state'' is an ever-growing transcript and the
transition is an unconstrained next-token distribution over it. Two
design moves convert this into a controlled POMDP. (i)~Event sourcing
with content hashing folds all relevant history into $s_t$, identified
by $h(s_t)$, rather than carried in an unbounded prompt.
(ii)~Validation as a transition guard prevents prose from silently
mutating state. Together $s_{t+1}$ becomes a function of $(s_t, a_t)$
alone---the kernel $F$ is literally Markovian, with the LLM relegated
to proposing $\Delta_t$ inside it.

\paragraph{Reservations.} Concurrent multi-NPC agency lifts this to a
parameterized-action Markov game~\cite{littman1994markov}, and we
leave that formal multi-agent analysis to future work.

\paragraph{Note on instantiation.} In our implementation, $a_t$ is not
chosen \emph{ex ante}: the world-agent (or a delegated sub-agent)
co-emits a narration $o$ and a structured JSON delta $\Delta$ in a
single LLM call, and PDVA reads the realised $(k, x_k)$ from $\Delta$'s
schema. The PA-POMDP abstraction holds at the interface
$(\mathcal{S}, \mathcal{A}, \Omega, O, F, \rho_0)$; in practice $\pi$,
$F$, and $O$ share LLM calls---narrative and action are entangled
outputs.

\subsection{Worked Example: A Single Turn}

To make the loop concrete, we walk through one turn in a scenario where
a player moves their character into a tavern. The world state $S_t$
includes a location file, two active NPCs (the barkeep and a guard),
and the player's current position.

\emph{Plan.} The context builder reads the player's location
(\texttt{town\_square}) and gathers the two NPCs whose perception
radius intersects that location. The bounded context $p$ is a JSON
object containing the relevant state slice and the player's input:
\begin{samepage}
\begin{lstlisting}[language=]
{ "player_input": "I walk into the Rusty Anchor tavern.",
  "context": { "player": {"position": "town_square", "gold": 15},
               "nearby_npcs": ["barkeep_milo", "guard_ivor"],
               "location": "town_square" } }
\end{lstlisting}
\end{samepage}

\emph{Diff.} The world-agent model returns a response $r$ containing a
prose narrative for the player and a structured proposed mutation:
\begin{lstlisting}[language=]
{
  "narrative": "You push open the heavy oak door. The smell
    of ale and woodsmoke hits you. Milo the barkeep looks
    up from wiping a glass.",
  "delta": [
    {"path": "player.position", "op": "set",
     "value": "rusty_anchor_tavern", "actor": "player"}
  ]
}
\end{lstlisting}

In the form of \S3.2, this is the parameterized action
$a_t=(\texttt{move},\ \{\texttt{position}: \texttt{rusty\_anchor\_tavern}\})$.

\emph{Validate.} Each $\delta_j$ is checked: the path
\texttt{player.position} conforms to the entity schema
(SchemaOK), the proposing actor \texttt{player} has write permission on
their own position (PermOK), and walking into a public tavern requires
no skill check under the game rules (RuleOK). All three conditions
pass; the mutation is admitted.

\emph{Apply.} The file at \texttt{player.position} is updated from
\texttt{town\_square} to \texttt{rusty\_anchor\_tavern}. The
world-agent journals event
$e_t = (t, \text{player}, \text{move}, h(S_t), h(S_{t+1}),
\Delta^{\mathrm{applied}}_t)$ and appends it to the event log. The new
state $S_{t+1}$ is hashed; the commit is canonical.

This trace is what a reviewer can audit, a creator can debug, and a
player can branch from: every decision the engine made is visible in
the event log, and every state is recoverable by replaying the journal
from genesis.

\section{The World-Agent}

\begin{figure}[!t]
\centering
\includegraphics[width=0.92\linewidth]{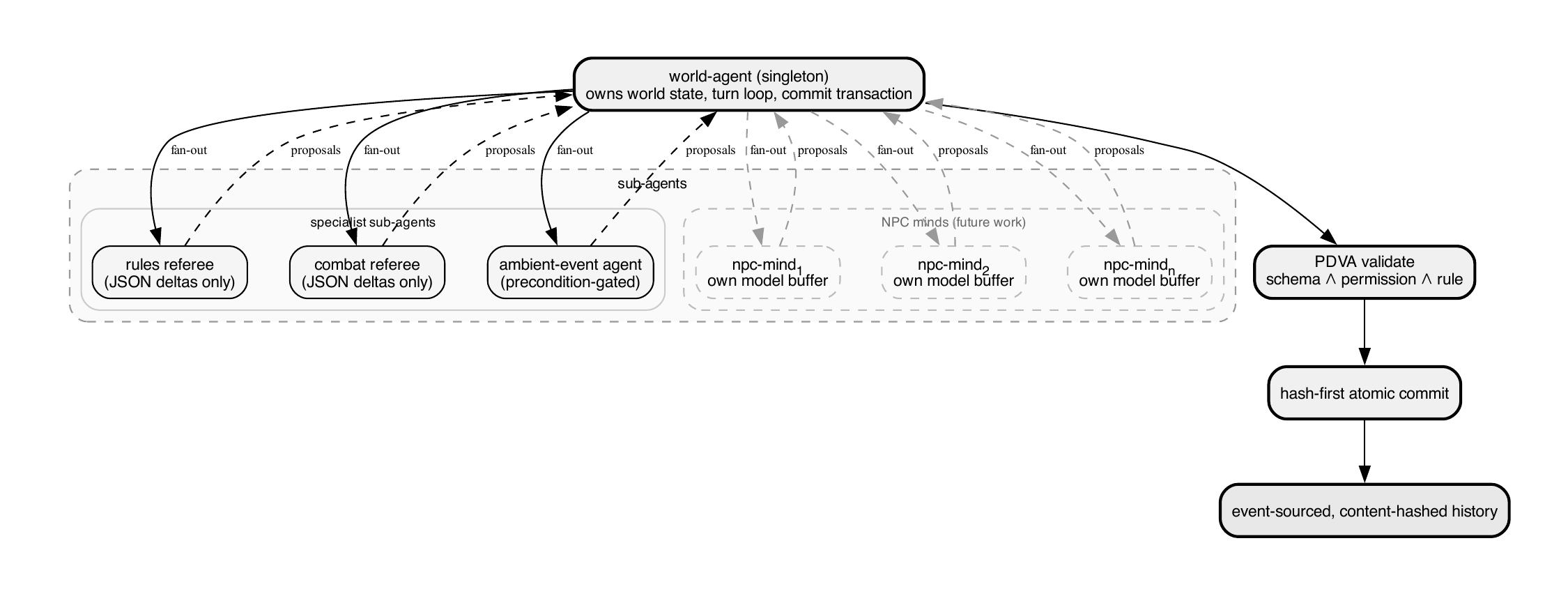}
\caption{The world-agent harness. A singleton world-agent (top) reads
the canonical JSON state and integrates proposals from specialist
sub-agents (rules, combat, ambient referees) and---in future
work---parallel NPC minds (dashed group) through the PDVA validation
stage, then commits one content-hashed update to an event-sourced
history.}
\Description{A vertical block diagram. A single world-agent box at the
top reads the canonical JSON world state. Below it, two grouped clusters
of sub-agents: specialist referees (rules, combat, ambient-event) and
NPC minds (shown dashed to indicate future-work parallelism). Solid
arrows show fan-out from the world-agent; dashed arrows show proposals
flowing back. Below the sub-agents, a PDVA validation box feeds a
hash-first atomic commit box, which produces the content-hashed,
event-sourced history at the bottom.}
\label{fig:harness}
\end{figure}

\S3 says \emph{how} the world advances; this section says \emph{who}
advances it (Figure~\ref{fig:harness}). Exactly one \textbf{world-agent}
per game owns the world directory, the turn loop, and the commit
transaction. It is the single writer per turn: any proposal---from the
player, a specialist sub-agent, or in future work an NPC mind---converges
through it into \emph{one} atomic, content-hashed commit, so there is no
race on world state. It also runs the world's own clock and ambient
dynamics (time progression, scheduled and precondition-gated events).

\textbf{Memory is injected, not recalled.} Each turn the world-agent
assembles a \emph{context pack}---a fixed seed set (player, scene frame,
active entities) plus relevance-ranked records under a small cap---and
injects those canonical JSON entries directly into the acting model's
prompt. The acting model's knowledge is thus a deterministic function of
committed world state, not of whatever survived in a chat
window---persistence is enforced by re-injection, not hoped for from
context length.

\textbf{The boundary is the contribution.} An agent contract
$c=(\mathrm{id},\mathrm{type},R,W)$ with read/write globs $R,W$ and a
deny-first default makes ``the world owns the world'' a checked
invariant rather than a prompt request. The end-of-turn collection is
\emph{hash-first}: the world-agent gathers each proposal, composes one
world commit, derives its identity by hashing content, and only then
writes to disk; if any write fails, nothing lands and no commit id is
recorded. Lifting this to multiple concurrent \textbf{NPC minds}---each
on its own model instance under a per-soul write subtree---is a
parameterized-action Markov-game extension we leave to future work
(\S6).

\begin{figure}[t]
\centering
\includegraphics[width=0.92\linewidth]{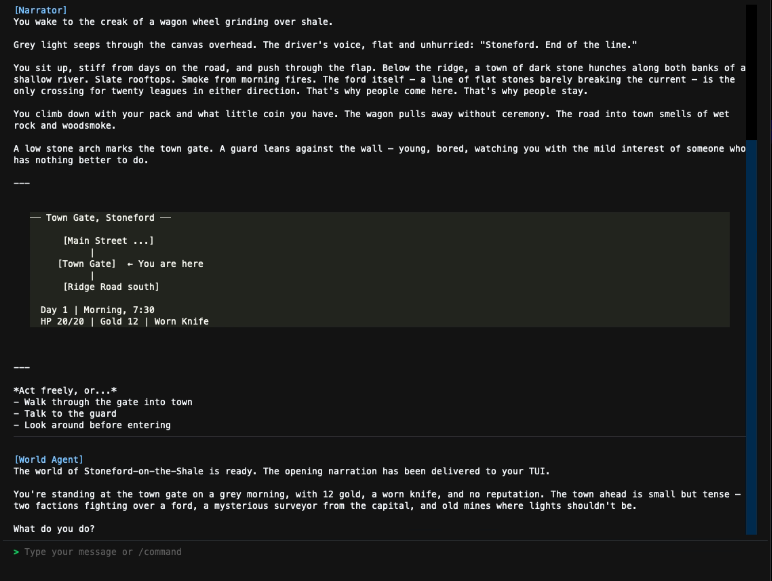}
\caption{The persistent loop as the player sees it. The world-agent
narrates (top), but the turn is anchored to durable, inspectable world
state---location, time of day, and the player's HP, gold, and
inventory---and the offered choices are a function of that state. The
\textit{[World Agent]} block reports the committed world: the player acts
into a world that will remember this turn.}
\Description{A text-interface screenshot of a role-playing session: a
narrator paragraph, a small ASCII map of a town gate, a status line showing
day, time, HP, gold, and a weapon, three numbered player choices, and a
world-agent status block confirming the committed world state.}
\label{fig:interface}
\end{figure}

\section{Early Observations}

Informal use by a small set of testers (N$\sim$5, three months) suggests
three hypotheses the planned study (\S6.3) will formally test: H1,
precondition-gated ambient events read as ``the world is alive'' even
when no character speaks; H2, narrated validation rejections
\emph{increase} felt consequence rather than break immersion; H3,
inspectable, hash-identified state lowers the barrier to authoring and
debugging worlds.

To move past impressions, we mined the engine's own play history---session
traces, narrated playthroughs, and the event log---for incidents
illustrating the three mechanisms in action, yielding 15 distinct
episodes (P1:~9, P2:~5, P3:~1) that each cite a source file and turn
range; the catalogue exemplifies the framework as a real deployment,
not as a substitute for the planned player study below. \textbf{(P1)~Multi-agent}: the harness
fans each turn out to specialist and character sub-agents and reconfigures
which are live as the scene moves between locations.
\textbf{(P2)~JSON-forced memory injection}: a named entity resurfaces
sixteen turns after it was introduced, and a character-agent consolidates
an early encounter into long-term memory with a declared behavioural
impact (``the first companion I could trust'')---recall driven by
re-injected state, not chat history. \textbf{(P3)~JSON state flow}: each
turn is a schema-validated delta applied to JSON files and content-hashed.
Because state lives in JSON, the catalogue is auditable rather than
anecdotal.

\section{Open Problems and Planned Evaluation}

\subsection{Open Problems}

\textbf{Determinism under LLM stochasticity.} Hash-first commits make
\emph{state} replayable, but reproducing a run requires recording every
model call (request and response) so replay can play them back instead
of re-sampling. Designing this record/replay interface without
ballooning storage is open.

\textbf{Per-actor cost discipline.} Once concurrent NPC minds land,
one model instance per actor multiplies API spend; per-turn token and
wall-clock caps and a concurrent-actor cap are necessary, and their
effect on pacing and perceived character ``aliveness'' is unstudied.

\textbf{Narrative drift on under-specified detail.} PDVA validates
only what is in the schema; fine-grained NPC traits not tracked as
canonical state (e.g.\ which hand carries a scar, the tone of a voice)
can drift across turns---we observe an NPC's missing finger swapping
between left and right hand across a 24-turn session. Closing this
gap requires either richer entity schemas or a downstream consistency
guard on narration $o_t$.

\textbf{Validity threats.} The Markov property we engineer applies to
world state $s_t$ via PDVA validation; rule coverage $\mathcal{R}$ is
necessarily incomplete and the prose observation $o_t$ is not itself
validated. The framework is tested primarily with one LLM provider;
whether the PA-POMDP properties hold across model families is
unverified. An artifact bundle (code, scenario assets, study
instruments) accompanies the project, WorldLines. The player study
below is planned rather than executed, so all player-experience claims
remain open.

\paragraph{Note on apparatus.} A pilot apparatus---a no-engine
single-LLM baseline, the same LLM with canonical JSON injected each
turn, and the full multi-agent loop, scored by an LLM judge---exists
and is released with the artifact, but at the scales we have tried
($n\!=\!1$--$3$ scenes, a few turns, single runs) judge variance
dominates and our recall construct conflates \emph{non-contradiction}
with \emph{recall}, so we report no effect size. Effect-size estimation
is deferred to the powered study below.

\subsection{Planned Player Study}

We are designing a within-subjects experiment to compare stateless
chat role-playing (Condition~A: a single LLM with a persona prompt, no
persistent world state) against the persistent loop (Condition~B: the
world-agent/character-agent architecture described above), both
running the same LLM backend and the same scenario.

\emph{Design.} Within-subjects, 2 (condition) $\times$ 3 (session)
counterbalanced. Half the participants start with Condition~A across
three 20-minute sessions, then after a $\ge$24-hour washout, complete
Condition~B across three sessions; the other half receives the reverse
order. Each session is followed by a short questionnaire battery. The
final session of each condition concludes with a 15--20 minute
semi-structured interview.

\emph{Measures.} Our primary outcome is player experience measured by
the 11-item miniPXI (Player Experience Inventory short
form)~\cite{haider2022minipxi}, chosen over GEQ for its validated
factor structure and lower fatigue in repeated-measures designs. Our
secondary outcome is character attachment, measured by the
Player--Avatar Interaction (PAX) scale~\cite{banksbowman2016pax}. We also
collect a single-item willingness-to-return rating and a short
open-ended reflection per session.

\emph{Research questions.} RQ1: Does persistent world state increase
player experience (immersion, autonomy, meaning) compared to stateless
chat? RQ2: Does persistence increase character attachment? RQ3
(exploratory, qualitative): How does validated persistence change the
player's mental model of the AI world---from ``tool I converse with''
to ``world I inhabit''?

\emph{Analysis.} Linear mixed-effects models (condition and session
fixed, participant random) for the quantitative outcomes; reflexive
thematic analysis~\cite{braun2006thematic, braun2019reflexive} of the
interviews; H1--H3 from \S5 tested as planned contrasts.

\emph{Feedback sought.} As an early-stage study, we bring three specific
questions to the games and HCI research community: (a)~Are the measures (miniPXI and
PAX) the right instruments for this comparison, or should we
supplement them with behavioral telemetry from the event log?
(b)~Does the within-subjects design risk carryover effects that
counterbalancing cannot fully address? (c)~What dimensions of the
player's mental model should the interview protocol prioritize?

\section{Conclusion}

We framed an LLM-driven game world for a human player as a
\emph{Parameterized-Action POMDP} whose state is a tree of canonical
JSON entities and whose transition kernel is the LLM-driven PDVA
pipeline. The contribution is the world model:
an \emph{orchestrated reality} in which the world is the canonical,
audited object, and any acting model---a player, an NPC mind, or an RL
policy---acts on it through validated, parameterized actions $(k, x_k)$
and observes only $O(s)$. We seek discussion on the framing and the
path toward multi-NPC Markov games and learnable RL environments for
game world models.

\bibliographystyle{ACM-Reference-Format}
\bibliography{refs}

\end{document}